\documentclass[parallelisme]{compas2025}
\usepackage{cite}
\usepackage{amsmath,amssymb,amsfonts}
\usepackage{algorithmic}
\usepackage{graphicx}
\usepackage{textcomp}
\usepackage{xcolor}
\usepackage{url}
\def\BibTeX{{\rm B\kern-.05em{\sc i\kern-.025em b}\kern-.08em
    T\kern-.1667em\lower.7ex\hbox{E}\kern-.125emX}}

\toappear{1} 

\begin{document}

\title{Bridding OT and PaaS in Edge-to-Cloud Continuum}
\shorttitle{OTPaaS Concept}

\author{Carlos J. BARRIOS H. (1,2,3) and Yves DENNEULIN (1,2)}%

\address{LIG/INRIA, Datamove Team, Grenoble, France \\
Université Grenoble-Alpes, Grenoble-INP, Grenoble, France\\
Universidad Industrial de Santander, Bucaramanga, Colombia\\
carlos-jaime.barrios-hernandez@inria.fr, Yves.Denneulin@grenoble-inp.fr
}


\date{\today}

\maketitle

\begin{abstract}
\textbf{The Operational Technology Platform as a Service (OTPaaS)} initiative provides a structured framework for the efficient management and storage of data. It ensures excellent response times while improving security, reliability, data and technology sovereignty, robustness, and energy efficiency, which are crucial for industrial transformation and data sovereignty. This paper illustrates successful deployment, adaptable application management, and various integration components catering to Edge and Cloud environments. It leverages the advantages of the Platform as a Service model and highlights key challenges that have been addressed for specific use cases.
  \MotsCles{Platform as a Service, Edge-To-Cloud Continuum, Internet of Things.}
\end{abstract}

\section{Introduction}

Digital transformation drives organizations to enhance efficiency and reduce costs by integrating technologies within a hierarchical structure. This integration is vital for adapting to digital changes, allowing companies to seize opportunities \cite{b1}. In Industry 4.0, it merges physical and digital realms through big data, automation, and cloud computing \cite{b2}. Recognizing Operational Technology (OT) issues is crucial for effective monitoring. Key challenges include vulnerabilities, the need for data-driven decisions, and integrating diverse standards.

\begin{figure}[h!]
\includegraphics[scale=0.2]{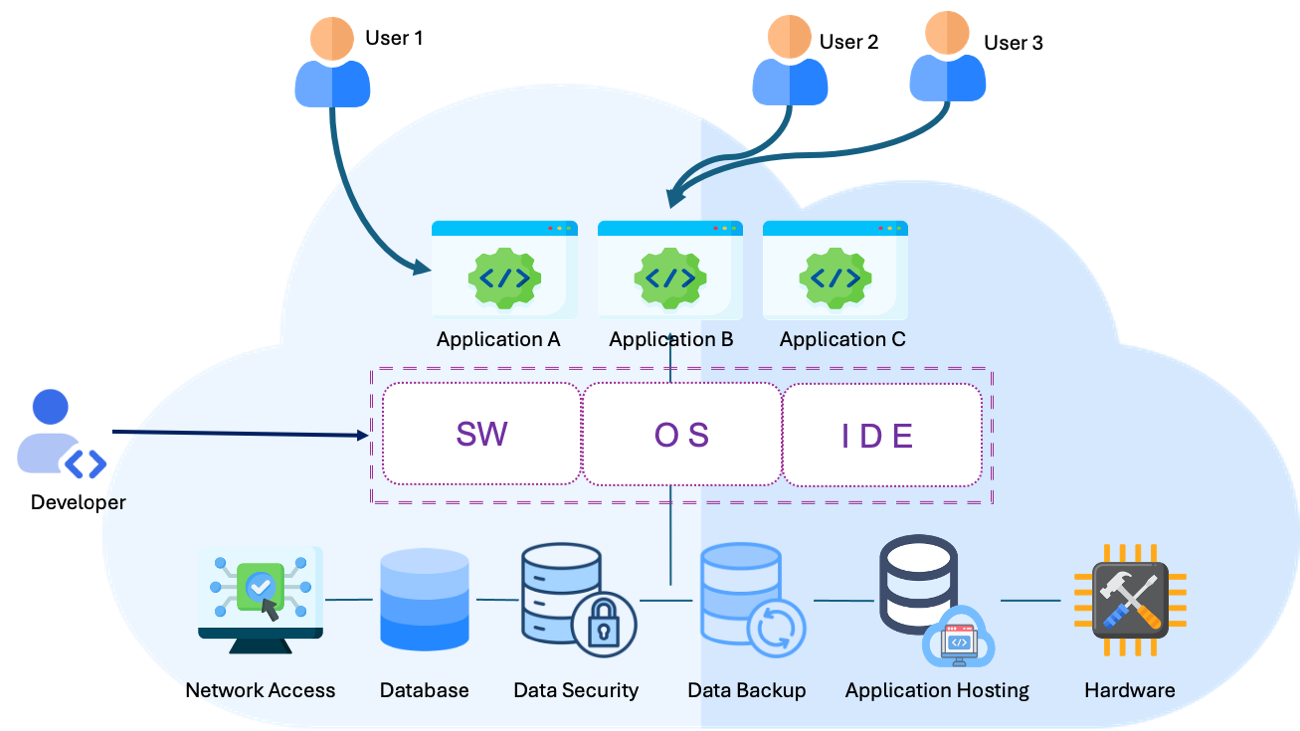}
\includegraphics[scale=0.08]{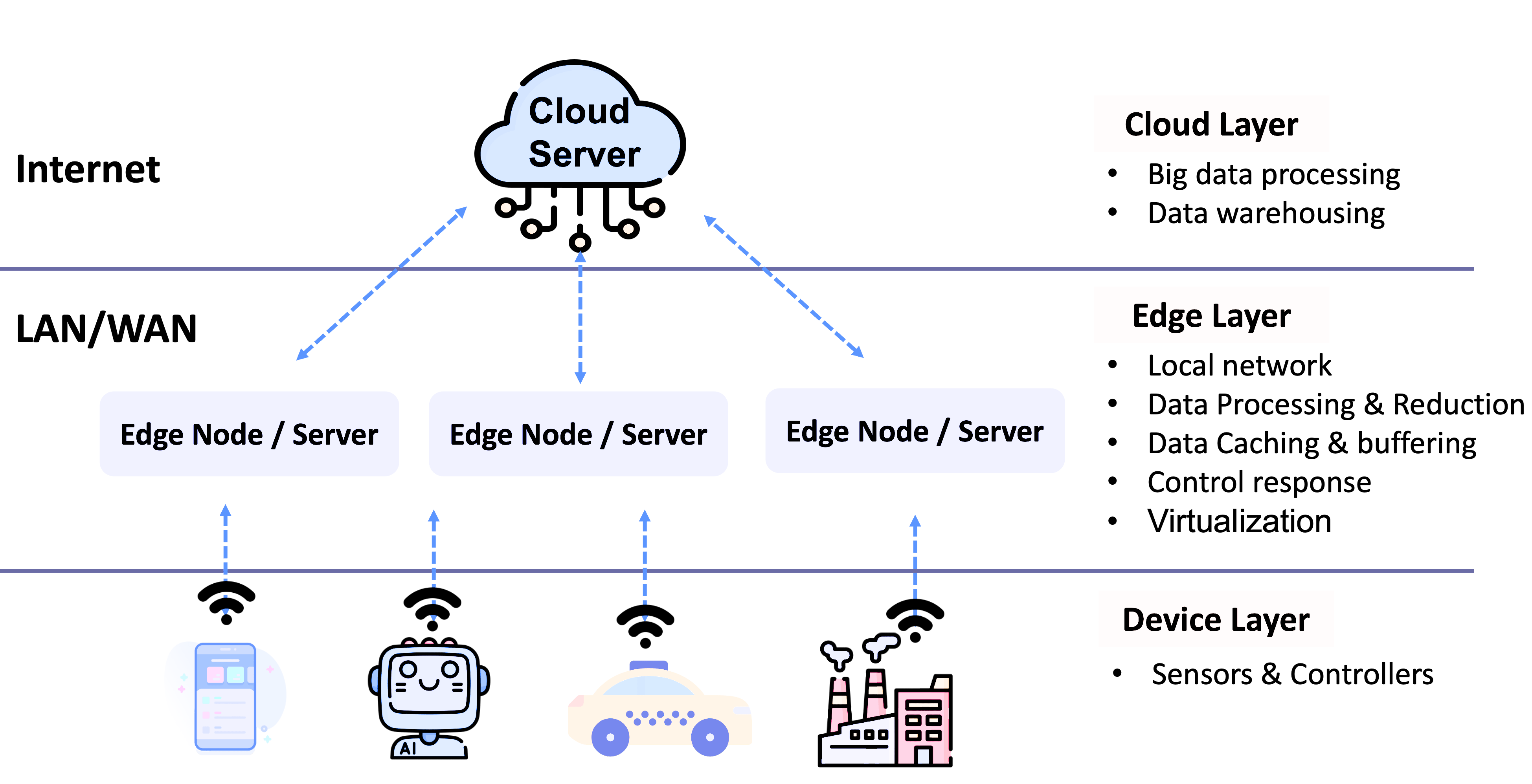}
\caption{Platform as a Service Workflow Model and Edge-to-Cloud Architecture}
\label{fig:PAAS}
\end{figure}

Integrating computing elements with industrial automation presents challenges. Although there are various technology methodologies, few are implemented successfully or are broadly applicable \cite{b3} \cite{b4}. This drive has led to advocacy for IoT initiatives aligned with Edge and Cloud Computing for industrial applications \cite{b5}\cite{b6} \cite{b7}. A Platform as a Service (PaaS) for Edge-to-Cloud Computing enhances frameworks and scalability, incorporating automation \cite{b8}. However, data security, compliance, and sovereignty issues persist. The OTPaaS initiative recommends a framework for effective data management within PaaS in the Edge-to-Cloud continuum, suggesting a platform for developing and managing applications while ensuring security and trust. A standard Platform as a Service (PaaS) includes infrastructure, cloud applications, and a Graphical Use Interface (GUI), similar to serverless computing or Function as a Service (FaaS) models \cite{b9}, where the cloud provider manages the infrastructure. Figure \ref{fig:PAAS} illustrates a PaaS workflow operation and Edge)To-Cloud Architecture. In this scenario, developers access various resources through specific network connections, which enable the application offerings for users. Users access applications and, in some cases, do so concurrently. Within this context, several key components are important to note: Integrated Development Environments (IDEs), Middleware (an invisible entity represented by dotted lines), Operating Systems (OS), Software (SW), Databases (including Data Security and Data Backup), along with Data and Application Hosting. A PaaS vendor is responsible for either managing servers, storage, and physical data centers, or acquiring these resources from a third-party provider. This arrangement enables development teams to focus their efforts on application development, rather than being concerned with infrastructure-related issues.

The left side of the Figure \ref{fig:PAAS} shows key points: the necessity of pre-configured environments for rapid development and deployment, easy collaboration access for distributed users, and operational support systems. To mitigate digital sovereignty concerns, OTPaaS introduces a continuum between the edge and the cloud, moving away from siloed data processing to the native \emph{Gaia-X} platform. \emph{GAIA-X}, launched by the European Union, sets a de facto standard for federated, trusted data and infrastructure ecosystems with its specifications, rules, policies, and verification framework \footnote{More information in: \url{https://gaia-x.eu}.}.

The Edge-to-Cloud computing Continuum marks a significant shift in data management across computing settings \cite{b10}. It integrates Edge computing, which processes data near sources like IoT devices, with centralized Cloud computing. This framework enables scalable, distributed operations tailored to specific needs, allowing applications to run from embedded devices to large infrastructures \cite{b11}, \cite{b12} \cite{b13}. Its integration is vital for real-time applications, such as smart manufacturing. Edge-to-Cloud architectures combine the strengths of both models, enhancing resource allocation and performance. The Edge layer facilitates immediate processing, while the Cloud offers storage and computational power as scalability demands. The right side of the Figure \ref{fig:PAAS} illustrates a generic Edge-to-Cloud architecture within a distributed environment, distributing processing across multiple Edges for data collection and analysis, connecting to the Cloud for workloads \cite{b14}. Edge nodes extend capabilities beyond local networks, enabling data processing, caching, control, and virtualization. At the lowest layer, IoT devices generate data.  Design characteristics include data replication, synchronization, caching, prefetching, load balancing, scalability, and security. In Edge-to-Cloud deployments, recognizing patterns poses challenges. Flexibility, storage sharing, and access, with QoS support, enhance integration and foster operational automation \cite{b15} \cite{b16}. 

This paper discusses how integrating OT and PaaS can successfully realize the benefits of an Edg2cloud computing continuum vision, combining the specific characteristics of both models.

\section{From Specificity To Open Cloud Solutions: The OTPAAS Concept}

The utilization of Open Cloud is not a recent development; it has been in existence for over a decade. Through various experiences, including those within the realm of video games, this strategy has been developed and evolved, as referenced in previous studies \cite{b17} \cite{b18}. The advantages of flexibility, cost effectiveness, and the ability to bypass vendor lock-in \cite{b19} have driven this evolution. The integration of solutions based on open-source technologies facilitates a diverse array of Cloud services tailored for operational technology, which encompasses Platform as a Service, thereby enabling the creation of a conceptual model for implementation and deployment, designated as OTPPaaS.

\begin{figure}[h!]
\centerline{\includegraphics[scale=0.25]{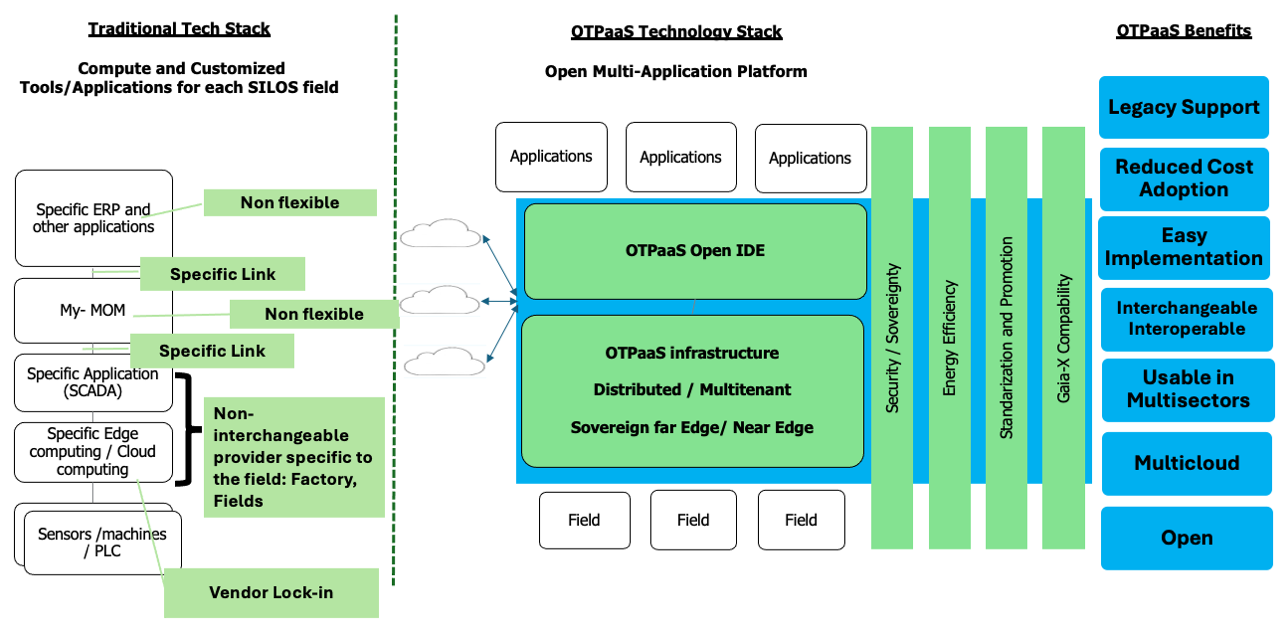}}
\caption{OTPaaS Concept}
\label{fig:OTPaaSConcept}
\end{figure}

Figure \ref{fig:OTPaaSConcept} illustrates the Open Cloud concept for OTPaaS, offering custom solutions in a multi-Cloud environment by converting field data processing into structured silos managed by specialized applications. Implementing the OTPaaS technology stack improves specialization and aligns with the Platform-as-a-Service model in Figure \ref{fig:PAAS}. A distributed, multitenant infrastructure allows application instances to serve multiple tenants, increasing resource efficiency. Challenges include development interfaces, data governance, costs, and energy savings. GAIA-X guarantees OTPaaS security, sovereignty, efficiency, and legacy support while promoting cost savings and flexible interoperability in an open-source multi-Cloud framework. OTPaaS fosters an ecosystem integrating Containers as a Service (CaaS) for application development, establishing a framework for all organizations. Tailored strategies are vital for flexibility and security, resulting in a hybrid implementation model.

\section{Efficient Operation Technology Integrating CaaS in PaaS}

\begin{figure}[h!]
\includegraphics[scale=0.35]{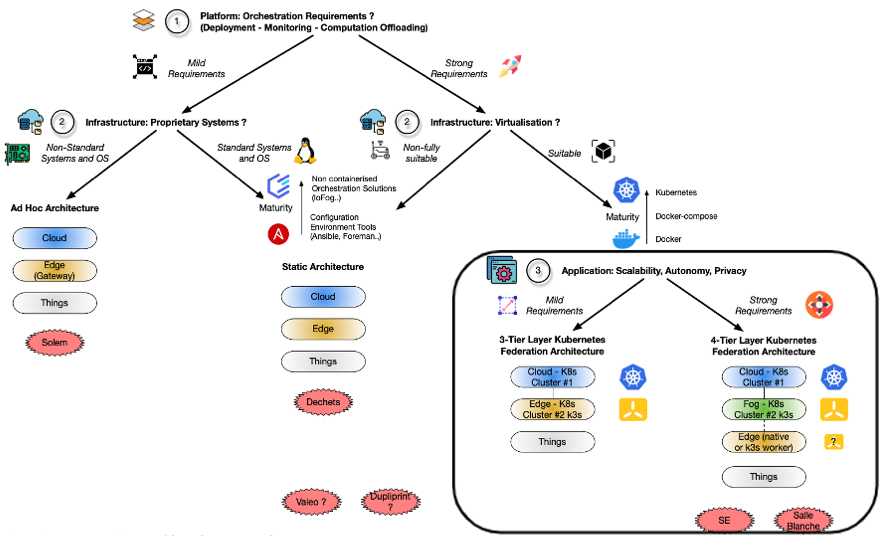}
\includegraphics[scale=0.35]{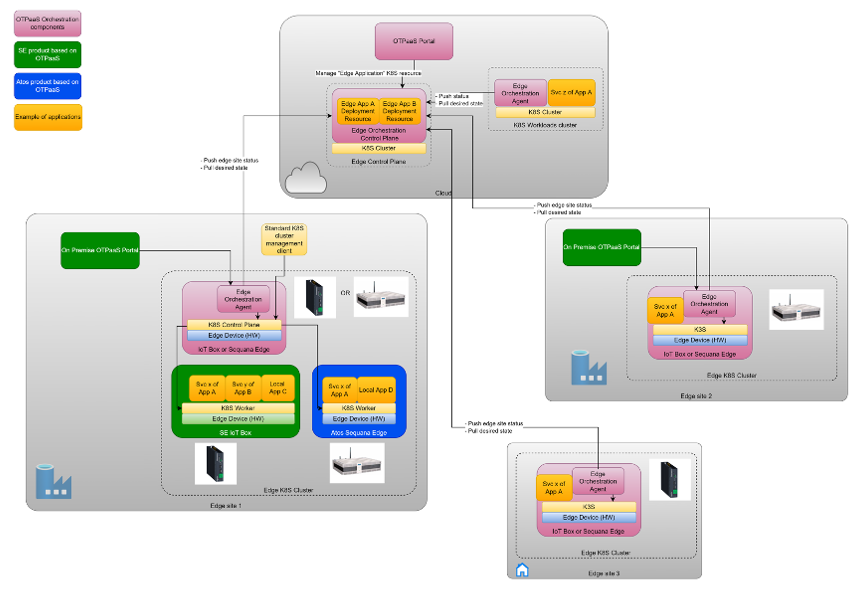}
\caption{PaaS Usage Requirements and OTPaaS Architecture}
\label{fig:OTPaaSNS}
\end{figure}

Orchestration requires deployment, monitoring, and computation offloading through on-premise or virtual setups. Figure \ref{fig:OTPaaSNS} illustrates that, despite different use cases, they share orchestration needs addressed by the OTPaaS architecture, which integrates CaaS with a modular approach for specific containerization. A single architecture is inadequate for OTPaaS, necessitating a flexible hybrid multi-cloud strategy. To ensure scalability, autonomy, and privacy, services use Kubernetes\footnote{\url{https://kubernetes.io/}}. Microservices permit customized development, and containers execute tasks independently. This fosters rapid prototyping and feedback prior to broader deployment. The multi-cloud strategy in OTPaaS streamlines management of containerized applications, enhancing OT and PaaS for portability and seamless operations.

OTPaaS architecture consists of orchestration, products (for organizations), and applications. Each module has specific containerization needs. Two require an on-premises OTPaaS portal, while one does not. They work through a shared resource pool via an Edge orchestration agent connected to the OTPaaS Cloud portal. This agent uses predefined states at the Edge site for efficient resource deployment. OTPaaS integration enables applications to adapt to fluctuating loads and resource demands, ensuring optimal performance. OTPaaS packages applications in containers for consistent performance across on-premises and cloud environments. CaaS platforms provide infrastructure to deploy, manage, and scale applications via APIs or GUIs. Developers utilize PaaS tools to build applications with databases, middleware, and services. OTPaaS automates orchestration and scaling for PaaS operations on CaaS platforms, ensuring high availability. The next section presents OTPaaS's mechanisms through three tailored use cases for different organizations and scales.

\section{OTPaaS Deployment and Uses Cases}

The test scales show effectiveness across a range of organizations, including large industrial IoT users, mid-sized scientific institutions, and small specialized companies. OTPaaS offers reliable, scalable, and secure hosted applications. This paper discusses challenges such as service integration, automated configurations, and resource allocation.

\subsection{IoT and Data Secure and Reliable}

Large organizations prioritize data security. A robust IoT system using a hybrid multi-cloud strategy combines on-premise, private, and hybrid clouds for optimal workload deployment while managing cost and compliance. The OTPaaS initiative requires a flexible, scalable, and resilient architecture with strong disaster recovery systems that ensure smooth data exchange and uninterrupted operations.

\begin{figure}[h!]
\centerline{\includegraphics[scale=0.6]{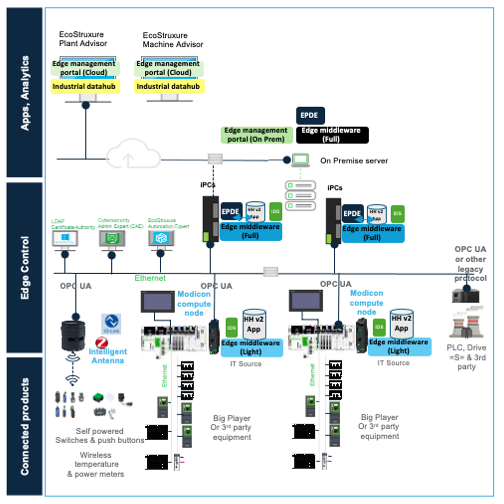}}
\caption{Use Case 1: Operational Architecture over OP CUA}
\label{fig:OTPPAASCase1}
\end{figure}

\begin{figure}[h!]
\includegraphics[scale=0.34]{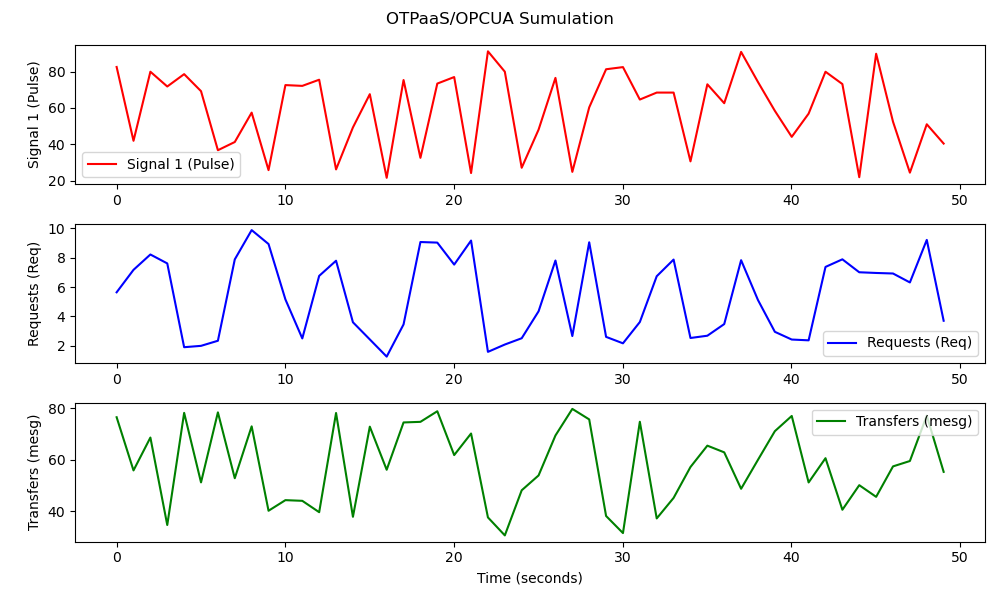}
\includegraphics[scale=0.46]{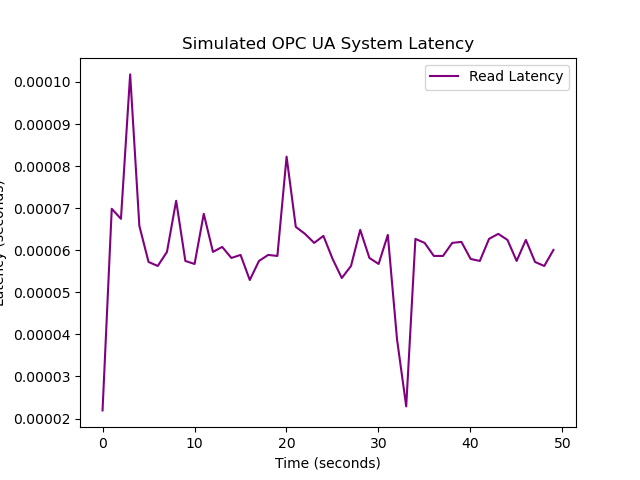}
\caption{OTPaaS Simulation on OPCUA}
\label{fig:OTPaaSOPCUA}
\end{figure}

Figure \ref{fig:OTPPAASCase1} shows the operational framework for the first use case, illustrating the OTPaaS concept from Figure \ref{fig:OTPaaSConcept} and the architecture in Figure \ref{fig:OTPaaSArchi}. This framework has three layers: the upper layer focuses on analytics, the middle manages edge control, and the lower includes device products. Schneider Electric's EcoStruxure Technology Platform integrates IoT and data security across various services\footnote{\url{https://www.se.com/ww/fr/work/campaign/innovation/overview.jsp }}. The SE IoT platform enables cloud management of connected devices. Application packages can be deployed through containers using the Edge runtime. In OTPaaS, improved edge nodes work with Kubernetes, allowing customization while managing resource integration, automated load balancing, and operational adjustments. The Unified Architecture of the Operations Platform (OPC-UA) enables automation systems to communicate and exchange data securely\cite{b20}. Its hierarchical node address space includes data variables, objects, methods, and components for machines, sensors, and controllers. OPC UA also supports diverse data types through object-oriented modeling, including complex structures and arrays. Based on the architectural abstraction from Figure \ref{fig:OTPPAASCase1}, we propose a 50-second simulation with 3 nodes for specific industrial use variables. Each node will simulate values (signals, requests, and message transfers) read by the client over time, as shown in Figure \ref{fig:OTPaaSOPCUA}.

We assess latency in a three-node simulation by measuring the time for a client to read server values. Figure \ref{fig:OTPaaSOPCUA} shows the results of the data retrieval per loop. Latency measures the time taken to send read requests and receive responses. The client connects to the server and tracks requests across nodes. In simulation, latency varies due to network delays and server processing. OPCUA provides cross-platform communication and scalability. The implementation advantages of the OTPAAS proposal are illustrated in the following use case.

\subsection{OTPaaS Efficient Orchestration}

Automation, coordination, and resource provisioning are vital for efficiency in Platform as a Service (PaaS). PaaS shifts Cloud tasks to the network's edge, bringing computational resources closer to data producers. This requires flexible and dynamic service orchestration, regardless of deployment methods like Function as a Service (FaaS) \cite{b21} or Infrastructure as Code (IaC) \cite{b22}. Implementing Cloud orchestration is crucial for managing diverse Cloud environments, automation workflows, integrating services, and optimizing resources The OTPaaS Platform features a centralized interface, the OTPaaS Portal, offering tools for constructing, deploying, and managing environments. Automate and streamline tasks on the platform to oversee the deployment, scaling, and lifecycle of services and applications.

\begin{figure}[h!]
\includegraphics[scale=0.17]{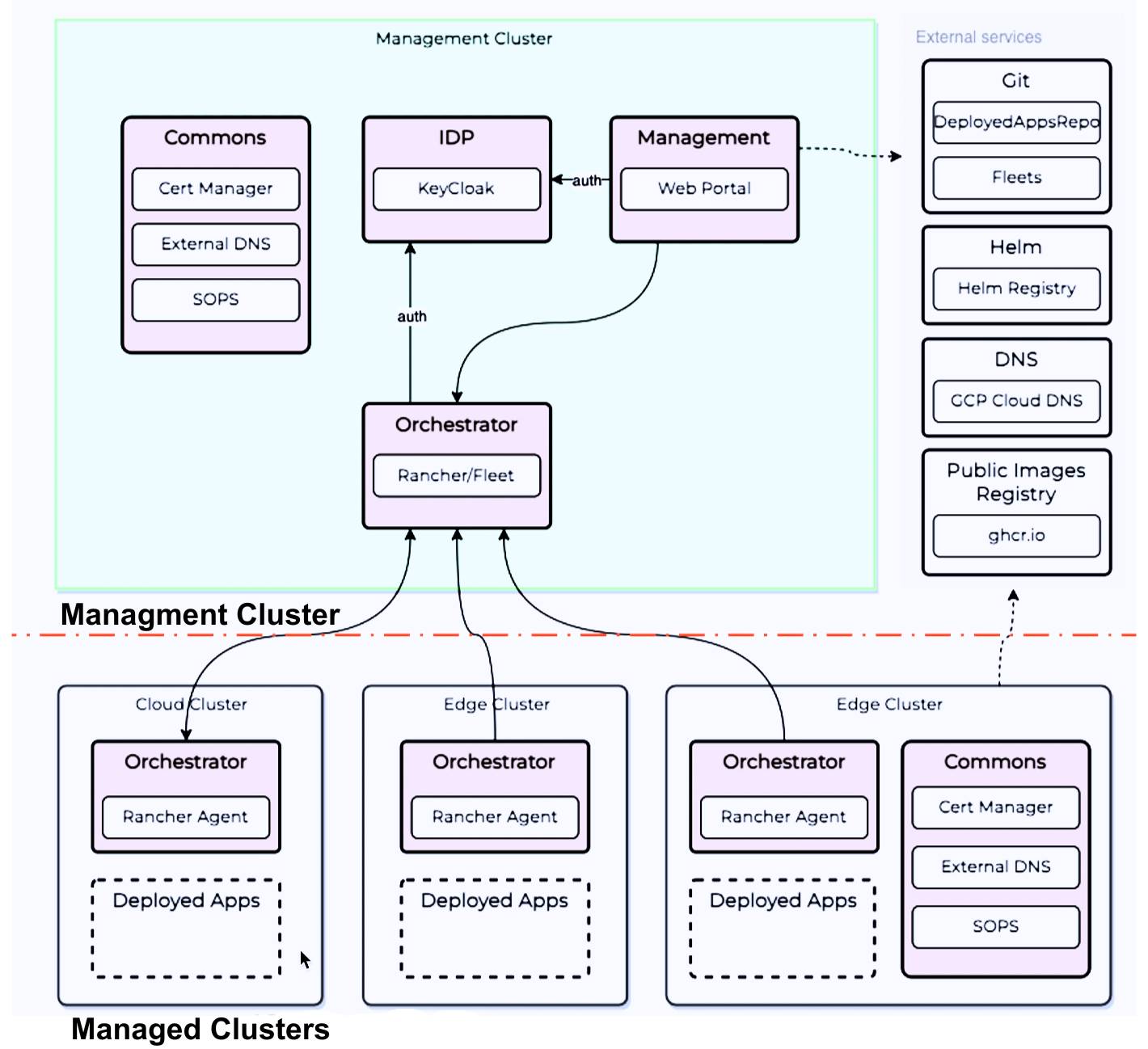}
\includegraphics[scale=0.24]{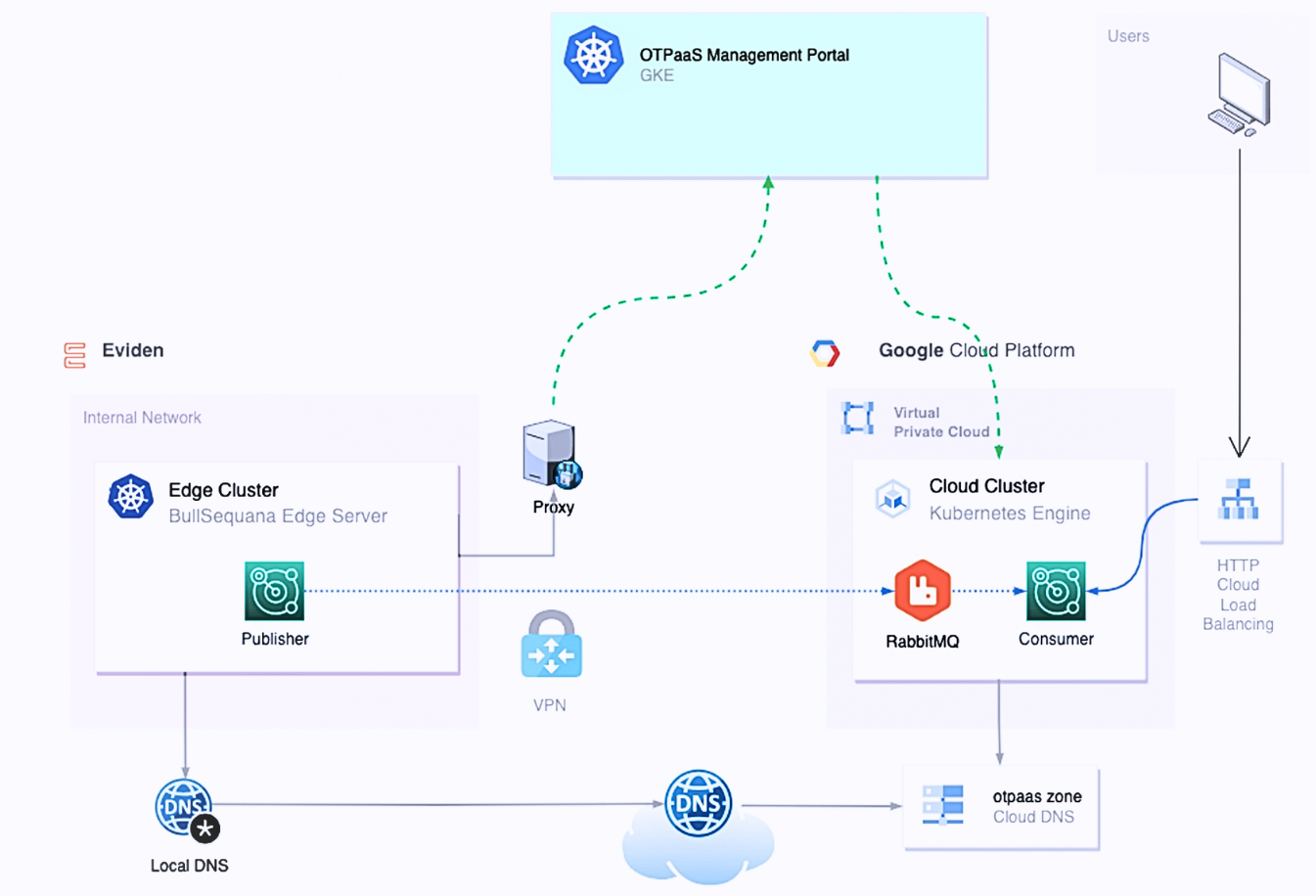}
\caption{Use Case 2: OTPaaS Orchestration and General Architecture }
\label{fig:ArchiInterne}
\end{figure}

Figure \ref{fig:ArchiInterne} shows the OTPaaS architecture for orchestration. The management cluster includes self-service components, a software catalog, lifecycle management (according to Continuum Cloud2Edge \cite{b23}), certification, authorization, and orchestration. The management, Cloud, and Edge clusters each have an orchestrator agent that manages local agents, facilitating application deployment and synchronization with the main orchestrator.

This use case focuses on automating application deployment, generating DNS entries and TLS certificates, applying customizations, and managing upgrades and removals. Messaging software such as RabbitMQ\footnote{\url{https://www.rabbitmq.com/}}, facilitates this by creating queues for applications to send messages. Figure \ref{fig:ArchiInterne} illustrates the OTPaaS architecture deployment using Eviden's use case developed by Eviden \footnote{\url{https://eviden.com/}}. This architecture enables the creation of Kubernetes clusters on Edge servers and in the Cloud. An agent, as depicted in figure \ref{fig:ArchiInterne}, synchronizes with the OTPaaS management portal. Communication between the agent and the Cloud platform is established through public access, while the Edge cluster agent can also synchronize with the portal. Depending on the task, support for actions may come from external or local services, such as Local or Cloud DNS.

\begin{figure}[h!]
\includegraphics[scale=0.4]{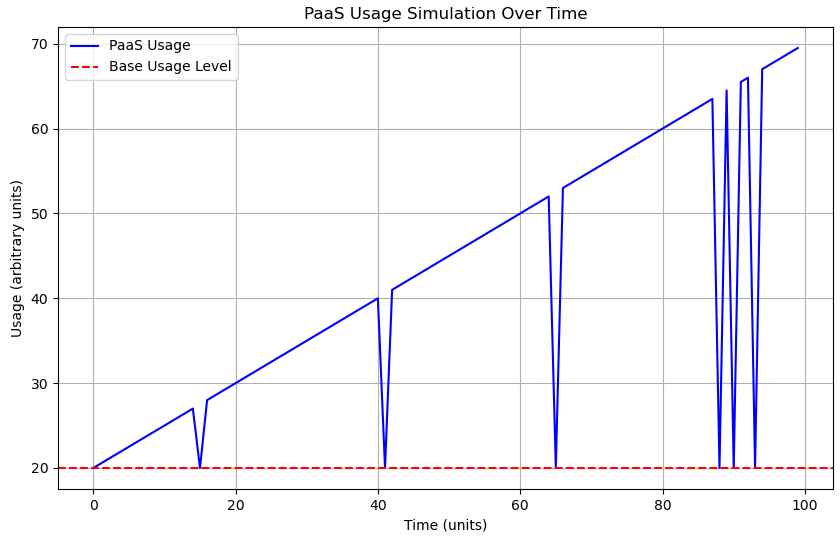}
\includegraphics[scale=0.28]{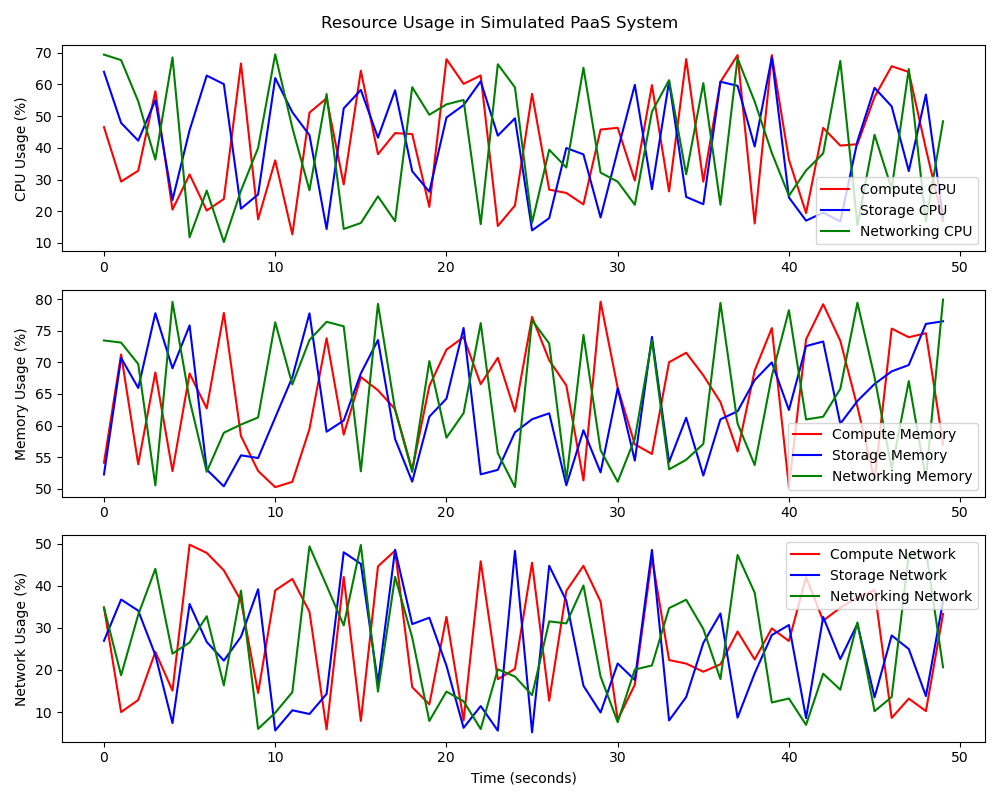}
\caption{OTTPaaS Simulated and Orchested Use}
\label{fig:OTPaaSusage}
\end{figure}

The platform serves as a foundation for monitoring the growth in utilization, focusing on its \emph{PaaS} characteristics. Figure \ref{fig:OTPaaSusage} illustrates a simulation of the use case, showing baseline usage (in red) and increasing utilization over time (in units), with periodic drops.

The simulation shows a minimal utilization of the platform of 20\%. Despite unique case characteristics and some downtime, usage increases significantly. Only 5\% to 10\% of the platform experiences downtime, yet usage continues to increase.

OTPaaS orchestration simulation is a system of services (nodes) coordinating with each other. The nodes may represent services like:

\begin{itemize}
    \item Compute Node: Running computational tasks.
    \item  Storage Node: Managing data storage and retrieval.
    \item Networking Node: Managing the communication between the other nodes.
\end{itemize}

This simulation models communication between services over time, representing task coordination across nodes while monitoring resource utilization (CPU, memory, network load) changes. Figure \ref{fig:OTPaaSusage} displays simulation results.


Three subplots will be generated as shown Figure \ref{fig:OTPaaSusage}: 
\begin{itemize}
\item CPU Usage: Shows the CPU utilization of the Compute, Storage, and Networking nodes over time.
\item Memory Usage: Shows the memory utilization of the nodes.
\item Network Usage: Displays the network usage for the nodes.
\end{itemize}

OTPaaS orchestration is efficient and ensures optimal resource allocation, flexible deployment, and seamless workload management. Developed tests demonstrate capacity to scale with workload demands. Automated scaling adjusts resources based on CPU or request load, enhancing orchestration while reducing operational overhead.

\subsection{Efficient Energy and Resource Allocation}

The OTPaaS concept boosts energy awareness and efficiency with dynamic scaling, workload optimization, and energy forecasting. It targets resource use optimization and energy-efficient scheduling, utilizing analytics to uncover power inefficiencies. The model handles complexities via feedback control loops for efficiency, scalability, and fault tolerance, benefiting multi-cloud and autonomic computing. A multi-loop architecture in Cloud computing uses feedback loops to optimize Cloud system management. This is crucial for autonomous computing, where systems monitor and adapt to changes, ensuring quality of service (QoS) and compliance with service level agreements (SLA) \cite{b24}.

\begin{figure}[h!]
\centerline{\includegraphics[scale=0.3]{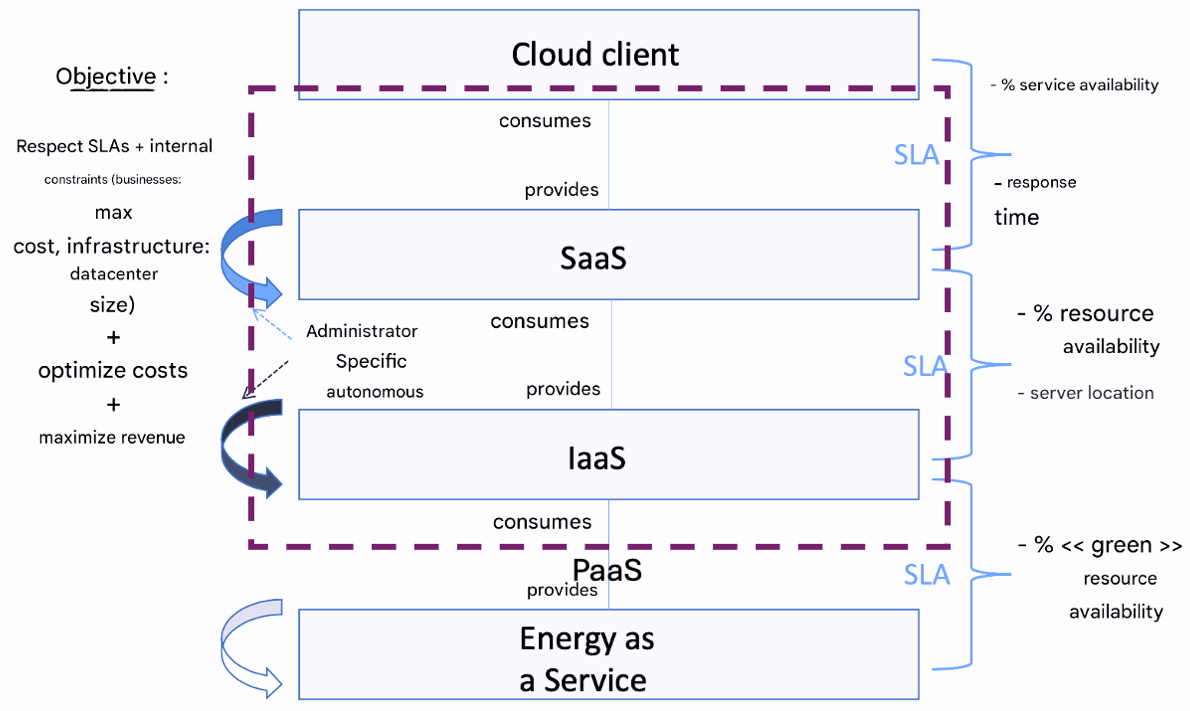}}
\caption{Energy as a Service Model in the Context of OTPaaS}
\label{fig:OTPaaSEaaS}
\end{figure}

Following the convergence of layers within the Edge-to-Cloud Continuum Model, OTPaaS is engineered to provide Energy-as-a-Service (EaaS) support, thus allowing access to energy-related services from designated energy resources to specialized service providers, contingent upon their availability, as depicted in Figure \ref{fig:OTPaaSEaaS}.

In OTPaaS, each control loop serves a specific purpose, such as resource allocation or workload balancing. The multi-control loop architecture synchronizes these loops to enhance Cloud performance and stability. Each operates uniquely: one feedback loop optimizes infrastructure, while others manage application performance. This effectiveness arises from a layered approach and collaboration among Edge devices, middleware, platform, and applications, with defined roles. Feedback loops ensure efficient operation and coordination across layers.

\begin{figure}[h!]
\centerline{\includegraphics[scale=0.5]{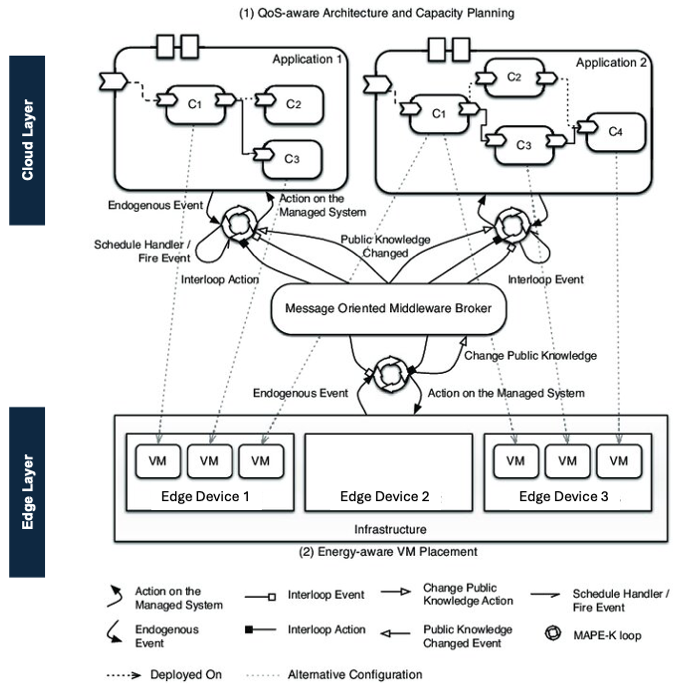}}
\caption{Use Case 3: Architecture Overview for the Self-adaptation Framework implemented in OTPaaS based in \cite{b25}}
\label{fig:Multiloops}
\end{figure}

\begin{figure}[h!]
\includegraphics[scale=0.26]{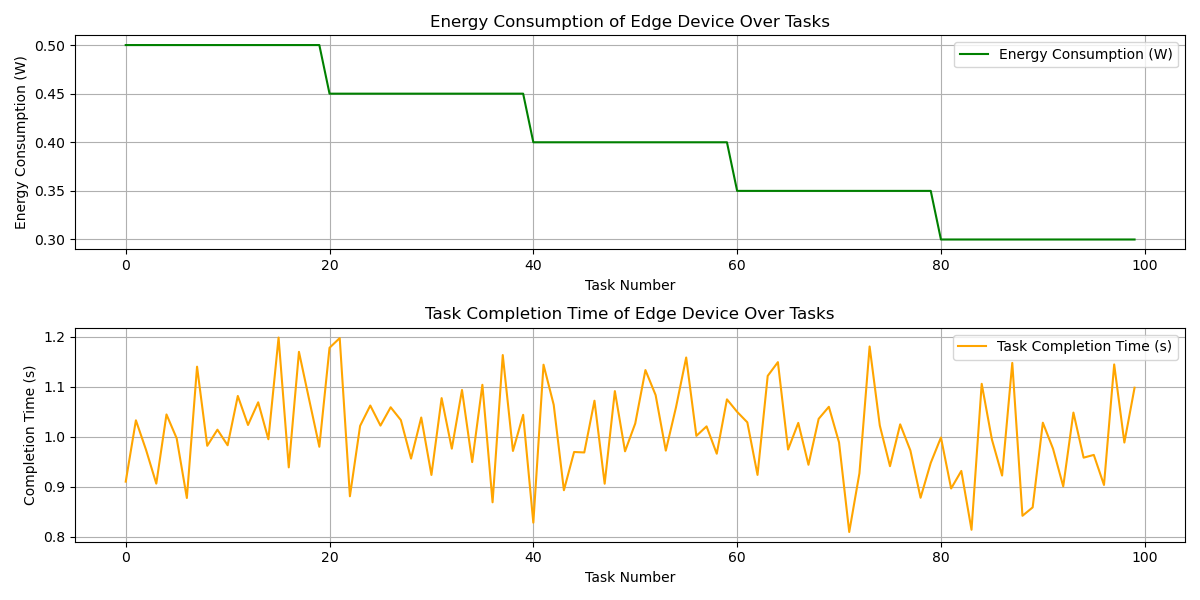}
\includegraphics[scale=0.26]{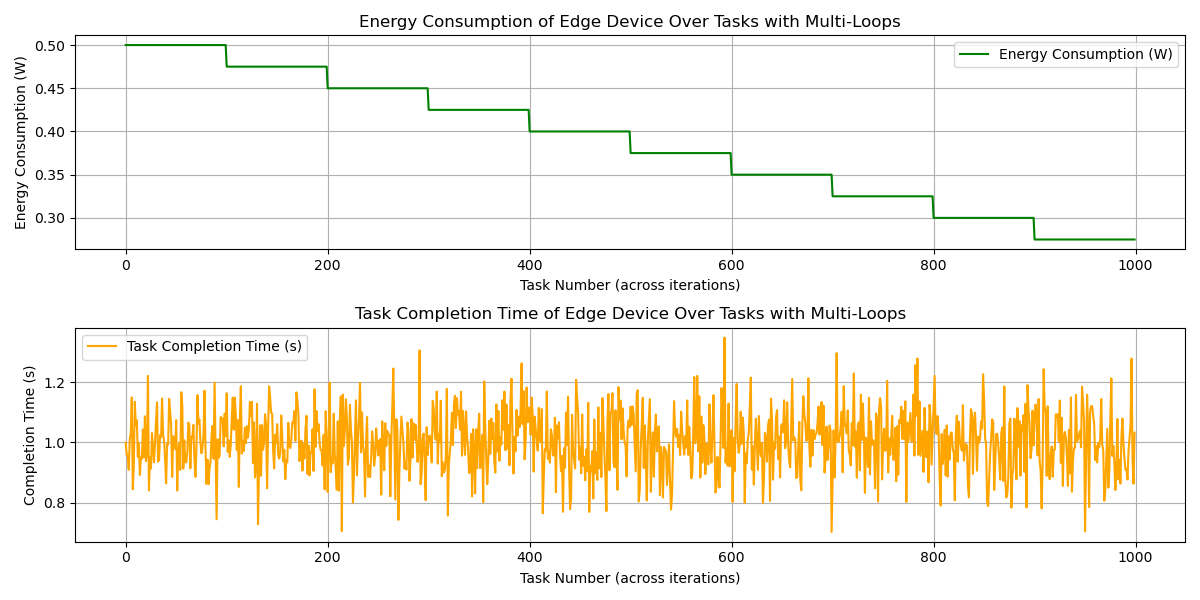}
\caption{OTPaaS Energy Consumption Behavior}
\label{fig:EnergySim1}
\end{figure}

The model proposed in \cite{b25} suggests that OTPaaS uses a framework of multiple autonomous managers (AMs) with the MAPE-K control loop \cite{b26} to allow self-adaptation for cloud applications and infrastructure. As depicted in Figure \ref{fig:Multiloops}, each application has a designated Application AM (AAM) to manage the elasticities of architectural and resource capacity. Identifies the optimal architectural configuration and the minimum required computing resources (VMs) to maximize Quality of Service (QoS) under specific workloads. At the infrastructure level, Infrastructure AM (IAM) oversees resources (Edge Devices and VMs) in an energy-efficient manner, optimizing VM placement to maximize their number across the fewest Edge Devices. Figure \ref{fig:EnergySim1} shows the energy-efficient strategy in OTPaaS for Edge devices performing AI tasks, using several autonomic managers. We evaluate energy consumption per task with the managers' improvements. As tasks progress, energy consumption declines due to these enhancements. Analyzing this data lets us generate random task completion times following a normal distribution, examining performance variability.

The findings highlight an energy-efficient OTPaaS strategy for devices with autonomic managers, focusing on energy consumption and AI task completion time. These managers coordinate Edge and Cloud resources, determining whether to execute tasks on energy-limited edge devices or offload them to the cloud based on energy use and latency, while adapting to workload and energy availability. The analysis of multi-loops evaluates energy consumption and performance in Edge devices during AI tasks, as shown in Figure \ref{fig:EnergySim1}. Energy usage per task decreases over iterations due to improved efficiency. Performance variations from multiple loops reveal energy consumption and task completion times under the conditions explained above.


\section{Background}

Cloud Platforms as a Service are evolving in industrial settings. For over two decades, industries have utilized cloud-based technologies, with innovations like IoT, cyber-physical systems, big data, and AI driving productive environmental trends \cite{b27}. To leverage PaaS, the goal is to embrace interoperability and integrate hybrid components to manage applications across public and private clouds, as shown by Cloud4SOA \cite{b28} \cite{b29}. This model supports resource provision for cloud processes handling data from edge devices, expanding technology adoption into a new Edge-to-Cloud Continuum model. It introduces smart distributed applications and raises questions about system capabilities, self-adaptation in complex environments, safety, security, robustness, and sustainability \cite{b30}.

In industrial settings, Operations Technology (OT) and Platform as a Service (PaaS) along the Edge-to-Cloud Computing Continuum are essential for digital transformation, especially for data control and compliance. While this article doesn't discuss legal issues, data sovereignty is critical. It involves data location, compliance, and interoperability. To tackle this, OTPaaS partners are collaborating with GAIA-X, which aims to create a secure European data infrastructure promoting sovereignty and transparency. This project fosters a digital ecosystem for data sharing and reduces dependence on non-European cloud providers\cite{b31}.  GAIA-X projects, backed by European funding, demonstrate effective principles across various organizations. For instance, the EuProGigant\footnote{\url{https://euprogigant.com}} initiative is a GAIA-X lighthouse project promoting Industry 4.0 and digital transformation. European organizations like NextCloud\footnote{\url{https://nextCloud.com/}} have joined the GAIA-X association to address business needs while ensuring transparency and interoperability. GAIA-X establishes a federated and secure data infrastructure for Europe. Similar global initiatives include International Data Spaces (IDS)\footnote{\url{https://internationaldataspaces.org/}}, the European Open Science Cloud (EOSC)\cite{b32}, and OpenStack\footnote{\url{https://www.openstack.org/}}, known for its Infrastructure as a Service (IaaS) capabilities, supporting public and private cloud development.

\section{Conclusion}

OTPaaS offers numerous benefits, and the Computing Continuum has great potential for digital transformation, but industry-academia partnerships face challenges. Operations Technology (OT) and Platforms as a Service (PaaS) are distinct yet linked within modern IT and Cloud Computing. Their intersection involves managing platform operations for high availability, performance, consistency, and resilience. The three scenarios show how CaaS can automate operations within OTPaaS. Here, orchestration is the main challenge, aiding data management, applications, and physical systems. Various experiments highlight OT process automation, data insights, and scalability. However, this solution raises questions on service models, particularly Edge-to-Cloud sharing concerns and energy efficiency, which might shift perspectives on operation technology towards EaaS in the Cloud Continuum.

\section{Discussion and Further Work}
The OTTPaaS project is vital for implementing technological solutions for digital transformation and competitiveness in France and Europe. It methodologically fosters interaction among various sectors of the productive and scientific ecosystem, enhancing DevOps tasks and user experience. However, it also presents challenges regarding conclusions that guide future efforts and approaches to digital transformation. It necessitates defining widely used concepts like the Computing Continuum, computing efficiency, and sustainability.

Integrating Operation Technologies (OT) with Edge-to-Cloud computing as Platform as a Service (PaaS) involves more than automation, focusing primarily on data security and privacy within a computing Continuum. Although resource allocation has improved, it is crucial to address the implications of the legislation governing data sharing and privacy, marking a significant change in the implementation of new SLAs \cite{b34} \cite{b35}. Concerns regarding data sovereignty arise, particularly in Europe but also globally. GAIA-X offers a framework for this, with OTPaaS aiming to align with its principles, which can also apply to other global frameworks, indicating key developments in both research and production.

Key aspects include network reliability and latency. Edge devices like sensors often disconnect from the OTPaaS platform due to instability. Despite its promise, this model faces challenges needing solutions for resilience and reduced latency via local data processing \cite{b36} \cite{b37}. Edge-Cloud integration can cause delays in large-scale aggregation. Effective scalability and management are vital for a distributed network of Edge devices and Cloud resources, necessitating tools for consistent performance. The diverse architectures in computing Continuum systems with OTPaaS pose challenges for research and implementation, and the lack of standardization complicates Edge-to-Cloud deployment. This article highlights benefits while addressing unresolved issues, viewing ecosystems as fragmented or suggesting subdivisions for enhanced flexibility and interoperability \cite{b38} \cite{b39}. Resource constraints at the Edge restrict on-site computing, limiting application deployment and energy efficiency across workloads.

\section*{Acknowledgment}
The OTPaaS project connects end users with technology providers, including major companies such as Atos/Bull, Schneider Electric, and Valeo, as well as SMEs like Agileo Automation, Mydatamodels, Dupliprint, Solem, Prosyst, and TwinsHel (by Soben). Leading French research institutions, including CEA, INRIA, and IMT, support it with the involvement of the Captronic Consortium. This initiative is made possible by the France Relance program, managed by the French Ministry of Economy, Finance, and Digital Industrial Sovereignty and funded by the European Union.



\end{document}